\documentclass{aa}
\usepackage{txfonts,bm}
\usepackage{color}

\usepackage{graphics,graphicx}
\usepackage{natbib}
\begin{document}
\title{Static compression of porous dust aggregates}
\author{Akimasa Kataoka\inst{1,2} \and Hidekazu Tanaka \inst{3} \and Satoshi Okuzumi \inst{4,5}\and Koji Wada \inst{6}}
\institute{
{Department of Astronomical Science, School of Physical Sciences, Graduate University for Advanced Studies (SOKENDAI), Mitaka, Tokyo 181-8588, Japan\\
\email{akimasa.kataoka@nao.ac.jp}
\and
National Astronomical Observatory of Japan, Mitaka, Tokyo 181-8588, Japan
\and
Institute of Low Temperature Science, Hokkaido University, Kita, Sapporo 060-0819, Japan}\and
Department of Physics, Nagoya University, Nagoya, Aichi 464-8602, Japan
\and
Department of Earth and Planetary Sciences, Tokyo Institute of Technology, Meguro, Tokyo, 152-8551, Japan
\and
Planetary Exploration Research Center, Chiba Institute of Technology, Narashino, Chiba, 275-0016, Japan
}

 \abstract 
 {
 In protoplanetary disks, dust grains coagulate with each other and grow to form aggregates. 
 As these aggregates grow by coagulation, their filling factor $\phi$ decreases down to $\phi \ll 1$.
 However, comets, the remnants of these early planetesimals, have $\phi \sim 0.1$.
 Thus, static compression of porous dust aggregates is important in planetesimal formation.
 However, the static compression strength has been investigated only for relatively high density aggregates ($\phi > 0.1$).
} 
 { 
 We investigate and find the compression strength of highly porous aggregates ($\phi \ll 1$).
 } 
{
We perform three dimensional $N$-body simulations of aggregate compression with a particle-particle interaction model.
We introduce a new method of static compression: the periodic boundary condition is adopted and the boundaries move with low speed to get closer.
The dust aggregate is compressed uniformly and isotropically by themselves over the periodic boundaries.
} 
{
We empirically derive a formula of the compression strength of highly porous aggregates ($\phi \ll 1$).
We check the validity of the compression strength formula for wide ranges of numerical parameters, such as the size of initial aggregates, the boundary speed, the normal damping force, and material.
We also compare our results to the previous studies of static compression in the relatively high density region ($\phi > 0.1$) and confirm that our results consistently connect to those in the high density region.
The compression strength formula is also derived analytically.
}
 {}
%


   \maketitle

\section{Introduction}\label{sec:introduction}
Planetesimal formation is a key issue in planet formation in protoplanetary disks \citep{Hayashi85,WeidenschillingCuzzi93}.
However, collisional growth of dust from sub-micron sized dust to kilo-meter sized planetesimals is still unknown.

In the growth process, one of the most important but unresolved problems is the internal structure evolution of dust aggregates.
Dust internal structure is important in planetesimal formation because dynamics of dust aggregates in protoplanetary disks is determined by coupling between gas and dust, in other words, size and internal density of dust aggregates.
In the early stage of dust coagulation in protoplanetary disks, the collision energy of the aggregates is too low to cause collisional compression \citep{Blum04, Ormel07,Zsom10, Zsom11b, Okuzumi12}.
As a result, the internal mass density $\rho$ decreases down to $\rho < 1.0 {\rm ~g ~cm^{-3}}$.

Both theoretical and experimental studies have shown that mutual collisions lead dust aggregates to have their fractal dimension $D \sim 2$, which is so-called ballistic cluster-cluster aggregation (BCCA) \citep{Smirnov90,Meakin91, Kempf99, BlumWurm00, KrauseBlum04, PaszunDominik06}.
The dust aggregates would be gradually compacted or disrupted in coagulation because of the increase in impact energy.
Such compaction has been investigated with numerical $N$-body simulations considering particle-particle interactions \citep{DominikTielens97, Wada07,Wada08,Wada09, Suyama08,Suyama12,PaszunDominik08, PaszunDominik09, Seizinger12}.

In most of previous studies investigating dust growth in protoplanetary disks, dust grains have been assumed to have constant internal mass density for simplicity \citep{Nakagawa81, Tanaka05, Brauer08a, Birnstiel10a}.
However, dust porosity evolves during dust growth in protoplanetary disks in reality.
In recent dust coagulation calculations, porosity evolution has been considered based on experimental and theoretical results \citep{Ormel07, Okuzumi+09,Okuzumi12, Zsom11b}.
They also suggested that $\rho$ decreases as $\rho \ll 0.1{\rm g ~cm^{-3}}$.

In the most recent work, though dust grains have size distribution, the dominant coagulation mode has shown to be similar-size collisions of dust aggregates \citep{Okuzumi12}.
As a result, their fractal dimension is approximately equal to 2 and their internal mass density $\rho$ has shown to become $10^{-5}{\rm ~g ~cm^{-3}}$ (equivalent to be the filling factor $\phi = 10^{-5}$ for ice particles with a density of $1.0 {\rm ~g ~cm^{-3}}$).
Such fluffy dust aggregates are believed to become planetesimals.
Since comets in our solar system, which would be remnants of planetesimals, have their internal mass density of  $\sim 0.1{\rm ~g ~cm^{-3}}$ \citep{AHearn11}, dust aggregates must be compressed from $\rho \ll 0.1 {\rm ~g ~cm^{-3}}$ to $\rho \sim 0.1{\rm ~g ~cm^{-3}}$ in protoplanetary disks.

Compression at dust aggregate collisions has been investigated in previous studies.
When collisional impact energy exceeds the critical energy, dust aggregates are compacted by their collsion \citep[e.g.][]{DominikTielens97, Suyama08, Wada07,Wada08,Wada09}.
However, the collisional compression is not effective to compress dust aggregates planetesimals \citep{Okuzumi12}.

One of the other compression mechanisms in protoplanetary disks is static compression by disk gas or self-gravity.
The static compression strength of dust aggregates has been investigated both experimentally and numerically \citep{PaszunDominik08, Guttler09, Seizinger12}.
However, they examined only relatively compact aggregates with $\rho \gtrsim 0.1{\rm g ~cm^{-3}}$ because their initial aggregates are ballistic particle-cluster aggregation (BPCA) clusters.
Because $\rho$ decreases down to $\rho \ll 0.1{\rm g ~cm^{-3}}$ at least in the early stage of dust growth, we need to reveal the static compression strength with $\rho \ll 0.1{\rm g ~cm^{-3}}$.

In this work, we investigate static compression of highly porous aggregates with $\rho < 0.1{\rm g ~cm^{-3}}$ by means of numerical simulations and analytical approach.
It is challenging to perform numerical simulations of static and uniform compression of highly porous aggregates.
Because such porous aggregates have low sound speed, we have to compress them in much slower velocity than in the case of compact aggregates, as will be shown in our simulations.
Such a slow compression of the fluffy aggregates costs much computational time.

In previous numerical studies of static compression, a dust aggregate is compressed by a wall moving in one direction \citep{PaszunDominik08, Seizinger12}.
However, this method has disadvantages to reproduce uniform and isotropic compression.
There are also side walls which do not move.
Such side walls also obstruct the tangential motion of monomers in contact with the walls, causing artificial stress on the aggregate and restructures them.
Moreover, since they measure the pressure with the force on the moving wall, the side walls may affect the pressure measurement.
In the present work, we develop a new method to reproduce static compression.
Instead of the walls, we adopt periodic boundary condition and the boundaries are getting closer to each other.
With this slowly-moving periodic boundaries, the aggregate is compressed uniformly and naturally.
The periodic boundary condition also enables us to represent a much larger aggregate than that inside the computational region.
This saves the computational time remarkably.

This paper is organized as follows: we describe the model of our numerical simulations in Section \ref{sec:simulation}.
We show the results of our simulations and find the compression strength in Section \ref{sec:result}.
We confirm the obtained static compression strength formula analytically in Section \ref{sec:analytic}.
We present our conclusion in Section \ref{sec:summary}.

\section{Simulation Setting}\label{sec:simulation}
We perform three dimensional numerical simulations of compression of a dust aggregate consisting of a number of spherical monomers.
As the initial aggregate, we adopt a BCCA cluster.
We solve interactions between all monomers in contact in each time step.
Interactions between monomers in contact are formulated by \citet{DominikTielens97} and reformulated with using potential energies by \citet{Wada07}.
We use the interaction model proposed by \citet{Wada07} in this work.
We briefly summarize the particle interaction model and material constants (see \citet{Wada07} for details).
Moreover, we describe the additional damping force in normal direction and the simulation setting in this section.
In our simulations, the aggregate is gradually compressed by its copies over the moving periodic boundaries.
This is an appropriate method to simulate uniform and isotropic compression.
We also describe the boundary condition in this section.
Since we do not have walls to measure the pressure in the periodic boundary condition, we use a similar manner of pressure measurement in molecular dynamics simulations.
We also introduce the method of pressure measurement below.

\subsection{Interaction Model}
We calculate the direct interaction of each connection of particles, taking into account all mechanical interactions modeled by \citet{DominikTielens97} and \citet{Wada07}.
The material parameters are the monomer radius $r_{0}$, surface energy $\gamma$, Young's modulus $E$, Poisson's ratio $\nu$, and the material density $\rho_{0}$.
Table \ref{tab:material} lists the values of the material parameters for ice and silicate.
\begin{table*}
\begin{center}
\caption{Material parameters in our simulation}
\begin{tabular}{ccc}
\hline
Material & ice & silicate\\
                &        & (same as \citet{Seizinger12})\\
\hline
Monomer radius $r_{0}$ [$\mu$m] &  0.1 &  0.6\\
Surface energy $\gamma$ [mJ ${\rm m}^{-2}$] &  100 & 20\\
Young's modulus $E$ [GPa] &  7.0 & 2.65\\
Poisson's ratio $\nu$  &  0.25 & 0.17\\
Material density $\rho_{0}$ [g~${\rm cm}^{-3}$] &  1.0 & 2.65\\
critical rolling displacement $\xi_{\rm crit}$ [\AA] &  8 & 20 \\
\hline
\end{tabular}\label{tab:material}
\end{center}
\end{table*}

We perform $N$-body simulations with ice particles except for one case with silicate particles.
In protoplanetary disks, ice particles are the most dominant dust material beyond the snowline.
Moreover, computational time required for calculation of ice particles is less than that of silicate.
Thus, we adopt ice particles in the most simulations.
We also perform a silicate case to compare with a previous study \citep{Seizinger12}.

The critical displacement still has a discrepancy between theoretical ($\xi_{\rm crit}=2$ \AA) and experimental ($\xi_{\rm crit}=32$ \AA) studies \citep{DominikTielens97,Heim99}.
We adopt the same parameter of \citet{Wada11}, $\xi_{\rm crit}=8$ \AA~ as a typical length for ice particles, and  $\xi_{\rm crit}=20$ \AA~ for silicate particles to compare with \citet{Seizinger12}.

$\xi_{\rm crit}$ is related to strength of rolling motion.
The rolling motion between monomers is crucial in compression. 
The rolling energy $E_{\rm roll}$ is the energy required to rotate a particle around a connecting point by 90 $\degr$.
The rolling energy can be written as
\begin{equation}
\label{eq:Eroll}
E_{\rm roll} = 6 \pi^2 \gamma r_{0} \xi_{\rm crit}
\end{equation}
(see \citet{Wada07} for details).
In the case of ice monomers, for example, $E_{\rm roll}=4.37\times 10^{-9}$ erg for $\xi_{\rm crit}=8 {\rm \AA}$.

We use normalized unit of time in our simulations.
In case of ice particles, the normalized unit of time is 
\begin{equation}
t_{0}=0.95 \left(\frac{\rho_{0}^{1/2}r_{0}^{7/6}}{E^{1/3}\gamma^{1/6}}\right) = 1.93\times 10^{-10} ~{\rm s}, 
\end{equation}
which is a characteristic time and represents approximately the oscillation time of particles in contact at the critical collision velocity (see \citet{Wada07} for details).

\subsection{Damping force in normal direction}\label{sec:damp}
The normal force between two monomers is repulsive when the monomers are close or attractive when they are stretched out.
Thus, normal oscillations occur at each connection.
For realistic particles, such oscillations would dissipate due to viscoelasticity or hysteresis in the normal force \citep[e.g.][]{GreenwoodJohnson06, Tanaka12}.
For such damping of normal oscillation, we add an artificial normal damping force to the particle interaction model, following the previous studies \citep{Suyama08, PaszunDominik08, Seizinger12}.

Assuming that two particles in contact have their position vectors ${\bm x_1}$ and ${\bm x_2}$, respectively, the contact unit vector ${\bm n_{c}}$ is defined as
\begin{equation}
{\bm n_{c}}=\frac{{\bm x_1}-{\bm x_2}}{|{\bm x_1}-{\bm x_2}|}
\end{equation}
 (see Figure 2 in \citet{Wada07}).
We introduce a damping force between contact particles in normal direction, defined as
\begin{equation}
{\bm F}_{\rm damp}=-k_{\rm n}\frac{m_{0}}{t_{0}}{\bm n_{c}}\cdot{\bm v_{\rm r}},
\end{equation}
where $k_{n}$ is the damping coefficient in normal direction and $m_{0}$ is the monomer mass.
The adopted value of $k_{n}$ is an order of 0.01.
To show that the result is independent of the normal oscillation damping, we perform $N$-body simulations with the damping factor $k_{n}$ as a parameter.

The timescale of damping is
\begin{equation}
\tau_{\rm damp} \sim \frac{t_0}{k_{n}} \sim 10^{2} t_{0},
\end{equation}
for $k_{n}=0.01$, is much shorter than the simulation timescale, which is typically $\sim 10^7 t_{0}$.
We show that the obtained compression strength is independent of the artificial normal damping force in our simulations (see Section \ref{sec:result_damp}).

\subsection{Uniform Compression by Moving Boundaries}\label{sec:periodic}
\begin{figure}[htbp]
 \begin{center}
  \includegraphics[width=80mm]{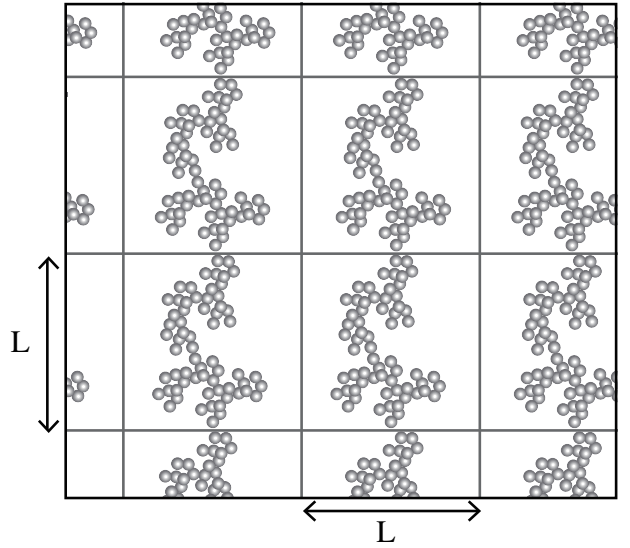}
 \end{center}
 \caption{
 Schematic drawing of the periodic boundary condition.
 Each of the box illustrates a boundary box with a side length $L$ for all direction.
 When the boundary starts to get closer, the aggregate sticks to the neighboring aggregates over the boundary and compressed by them.
 It should be noted that this picture is illustrated in 2D direction, but our simulations are performed in 3D.
 }
 \label{fig:2Dboundary}
\end{figure}

We adopt the periodic boundary condition in our simulations.
The aggregate in the computational region is surrounded by its copies, as shown in Figure \ref{fig:2Dboundary}.
Initially, we set a cubic box whose sides are periodic boundaries with a size of $L$ to be larger than the aggregate.
Thus, the initial BCCA cluster is detached from its neighboring copies over the periodic boundaries.
In our simulations, we gradually move the boundaries to the center of the aggregate to get closer to each other.
As a result, the aggregate sticks to the neighboring copies and is compressed by them in a natural way.
Therefore, the aggregate in the computational region corresponds to a small part of a whole large aggregate.
In other words, although the number of particles in numerical simulations are limited because of computational cost, the periodic boundary condition enables us to investigate a large aggregate, such as a $\sim$ cm-sized dust aggregate in protoplanetary disks.

Another advantage of the periodic boundary condition is that we do not need to introduce the wall for compression.
In the previous $N$-body simulations of static compression, dust aggregates are compressed by using the wall against the dust aggregate \citep{PaszunDominik08, Seizinger12}.
The wall itself may have some artificial effects on such experiments.
For example, the wall moves in one direction and thus this may be different from isotropic compression.
Besides, wall-particle interaction is different from particle-particle interaction, and thus it must be treated carefully.
In contrast, the periodic boundary condition does not need walls for compression because a dust aggregate is compressed by the neighboring aggregate over the periodic boundary.
In addition, the periodic boundaries in three directions make it possible to compress the aggregate isotropically.
Note that we calculate not only the interactions of particles in contact inside the computational region but also the interactions of the particles in contact across the periodic boundaries.
Thus, no special treatment of interactions, which is wall-particle interactions in the case of simulations with walls in previous studies, is required when a particle crosses the periodic boundaries.

The computational cubic region has length $L$ and the coordinates in $x, y$, and $z$ directions are set to be $-L/2<x<L/2$, $-L/2<y<L/2$, and $-L/2<z<L/2$, respectively.
We adopt periodic boundary conditions for all directions to reproduce a part of a large aggregate.
$L$ decreases with time $t$, $L=L(t)$.
The initial size of the box $L_{0}$ is adopted as the maximum size of the dust aggregate in $x$, $y$, and $z$ directions.

With the settings above, we move the boundaries of the computational region toward the center of the region.
The velocity at the boundary is given by
 \begin{equation}
v_{\rm b}=-\frac{C_{\rm v}}{t_{0}} L(t),
\label{eq:vb}
\end{equation}
where $C_{\rm v}$ is a dimensionless parameter (we call $C_{\rm v}$ the strain rate parameter hereafter).
Owing to this definition of the boundary speed, the aggregate is compressed at a constant strain rate independent of the region scale $L$.

The box size decreases with the constant rate $C_{\rm v}$ in three directions.
This corresponds to isotropic compression.
Since $\frac{dL}{dt}=2v_{b}$, the box size is written as
\begin{equation}
L=L_{0}\exp \left(-2C_{\rm v} \frac{t}{t_0} \right).
\end{equation}
Therefore, the whole time of compression is order of $t_{0}/C_{\rm v}$.
Typically we chose $C_{\rm v}=3\times10^{-7}$ and thus the compression time is $\sim 3\times10^{6}t_{0} \sim 0.6 {\rm ~ms}$.

\begin{figure}[htbp]
 \begin{center}
  \includegraphics[width=80mm]{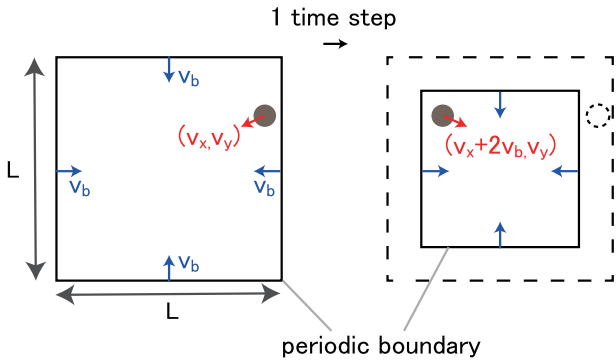}
 \end{center}
 \caption{
 Schematic drawing to illustrate how the particle velocity is calculated when a particle crosses a periodic boundary.
 For simplicity, we consider this situation in two dimensional field but we actually calculate this in three dimensional situation.
 We consider that a dust particle is close to the boundary in the left figure.
 In the next time step, the particle crosses the boundary (dashed circle in the right figure).
 We put the particle on the other side of the boundary as expressed in Equations (\ref{eq:mapx2}) and (\ref{eq:mapx1}).
 The velocity component is converted as expressed in Equations (\ref{eq:mapv2}) and (\ref{eq:mapv1}).
 This treatment well reproduces the isotropic compression in the velocity field.
 }
 \label{fig:gradv}
\end{figure}

When a particle crosses a periodic boundary, the velocity should be treated carefully to reproduce the quasi-static compression with periodic boundary condition.
Figure \ref{fig:gradv} illustrates how to calculate the velocity of particles across the periodic boundary.
When a particle goes out of the computational region across the boundary at $x=L/2$, we relocate the particle to the opposite side (i.e., from the boundary at $x=-L/2$).
In that case, the position of the particle in $x$ direction is converted as
\begin{equation}
x \longmapsto x-L
\label{eq:mapx2}
\end{equation}
Since the two boundaries at $x=-L/2$ and $x=L/2$ have a relative velocity of $2v_{\rm b}$, the $x$-component of the velocity $v_{x}$ of the particle is also converted as 
\begin{equation}
v_{x} \longmapsto v_{x}+2v_{b}.
\label{eq:mapv2}
\end{equation}
Owing to the conversion of $v_{x}$, the velocity of particle against the boundary which the particle crosses does not change before and after the crossing.
For a particle across the boundary at $x=-L/2$, the position and the velocity are converted as
\begin{equation}
x \longmapsto  x+L
\label{eq:mapx1}
\end{equation}
\begin{equation}
v_{x} \longmapsto v_{x}-2v_{b}.
\label{eq:mapv1}
\end{equation}
We also have the same treatments for particles across the boundaries at $y=\pm L/2$ and $z=\pm L/2$.

We introduce the constant strain rate at the boundaries for scaleless discussion.
However, the initial aggregate is not moving.
As the simulation starts, if all the particles in the aggregate are not moving, only the particles close to the boundaries have initial velocity.
This is not a constant strain rate.
To reproduce the scaleless constant strain rate initially, therefore, we initially give all monomers the velocity smoothly connected to the boundary speed.
The initial velocity is expressed as
\begin{equation}
{\bm v}({\bm r})=v_{\rm b}\times \frac{\bm r}{L_{0}/2},
\end{equation}
where ${\bm r}$ is the position vector of the monomers.

\subsection{Pressure Measurement}
In previous studies, a dust aggregate is enclosed by walls and the pressure was calculated by measuring the force exerted on the walls by the dust aggregate.
In this work, a dust aggregate is compressed by themselves because of the periodic boundary condition.
Therefore, we introduce another method to measure the pressure on the aggregate.
We calculate the pressure of the dust aggregate with the standard way in molecular dynamics simulations using the virial theorem as follows \citep[e.g.,][]{Haile92}.

Let us consider a virtual box that encloses an aggregate under consideration.
We define the force acting from the walls of the virtual box on the particle $i$ as ${\bm W}_{i}$, and the sum of the forces from other particles on the particle $i$ as ${\bm F}_{i}$.
The equation of motion of the particle $i$ is given by
\begin{equation}
\label{eq:eom}
m \frac{d^2 \bm{r}_i}{dt^2} = {\bm W}_{i} + {\bm F}_{i}.
\end{equation}
We take a scalar product of both sides of the equation with ${\bm r}_{i}$ and take a long time average of the both sides with time interval $\tau$.
The left-hand side becomes
\begin{equation}
m \frac{1}{\tau} \int^{\tau}_{0}{\bm r}_{i} \cdot \frac{d^2 \bm{r}_i}{dt^2} = m\frac{1}{\tau} \left[{\bm r}_{i} \cdot \frac{d \bm{r}_i}{dt}\right]^{\tau}_{0} - m\frac{1}{\tau}\int^{\tau}_{0}\frac{d \bm{r}_i}{dt} \cdot \frac{d \bm{r}_i}{dt} dt.
\end{equation}
The first term in the right-hand side vanishes in the limit of $\tau \to \infty$.
We define taking a long time average in $t$ as $\langle \rangle_{t}$
Taking summation of all particles and a long time average of Equation (\ref{eq:eom}), we obtain
\begin{equation}
\label{eq:eosave}
\left\langle \sum^{N}_{i=1} \frac{1}{2}m\left( \frac{d \bm{r}_i}{dt} \right)^2 \right\rangle_{t} = -\frac{1}{2} \left\langle \sum^{N}_{i=1} \bm{r}_{i} \cdot ({\bm W}_{i} + {\bm F}_{i}) \right\rangle_{t}.
\end{equation}
The first term in the right-hand side is related to pressure $P$.
The pressure is an average of all forces acting on the wall from all particles.
Using the normal vector ${\bm n}$ of the wall surface directed outward, the force received by the wall which has an area $dS$ is $P{\bm n}dS$.
Therefore, 
\begin{equation}
\left\langle \sum_{i}\bm{r}_{i} \cdot {\bm W}_{i}  \right\rangle_{t} = -\int_{S}P {\bm n} \cdot {\bm r}dS = - 3PV.
\end{equation}
This equation is obtained by taking surface integral as
\begin{equation}
\int_{S}{\bm n}\cdot{\bm r}dS=\int_{V}{\rm div} ~{\bm r} dV = \int_{V} \left( \frac{\partial x}{\partial x} +  \frac{\partial y}{\partial y} +  \frac{\partial z}{\partial z}\right)dV=3V.
\end{equation}
The translational kinetic energy $K$, averaged over a long time, is given by
\begin{equation}
K=\left\langle \sum^{N}_{i=1} \frac{1}{2}m\left( \frac{d \bm{r}_i}{dt} \right)^2 \right\rangle_{t}.
\end{equation}
Using $K$ and $P$, Equation (\ref{eq:eosave}) gives an expression of $P$ as
\begin{equation}
P=\frac{2}{3}K/V + \frac{1}{3}\left\langle\sum_{i} {\bm r}_{i}\cdot{\bm F}_{i} \right\rangle_{t}/V.
\end{equation}
We define the force from particle $j$ on particle $i$ as ${\bm f}_{i,j}$
The force ${\bm F}_{i}$ can be written as a summation of the force from another particle as
\begin{equation}
{\bm F}_{i}=\sum_{j\ne i}{\bm f}_{i,j}.
\end{equation}
Using ${\bm f}_{i,j}=-{\bm f}_{j,i}$, we finally obtain the pressure measuring formula as
\begin{equation}\label{eq:pressure}
P=\frac{2}{3}K/V + \frac{1}{3}\left\langle\sum_{i<j} ({\bm r}_{i}-{\bm r}_{j})\cdot{\bm f}_{i,j} \right\rangle_{t}/V.
\end{equation}
The first term in right-hand side of the equation represents the translational kinetic energy per unit volume and the second term represents the summation of the force acting at all connections per unit volume.
This expression is useful to measure the pressure of a dust aggregate under compression.
We do not need to put any artificial object such as walls in simulations because Equation (\ref{eq:pressure}) is totally expressed in terms of the summation of the physical quantities of each particle, which are the mass, the position, the velocity, and the force acting on the particle.
In our calculations, we take an average of pressure for every 10,000 time steps, correspondent to 1000 $t_{0}$ because we set 0.1 $t_{0}$ as one time step in our simulation.

As mentioned in Section \ref{sec:damp}, the adopted damping force corresponds to rapid damping of normal oscillations.
Thus, the kinetic energy of random motion rapidly dissipates.
This corresponds to the static compression and thus the compression strength is determined by the second term of Equation (\ref{eq:pressure}).

\section{Results}\label{sec:result}
\begin{figure*}[htbp]
 \begin{center}
  \includegraphics[width=180mm]{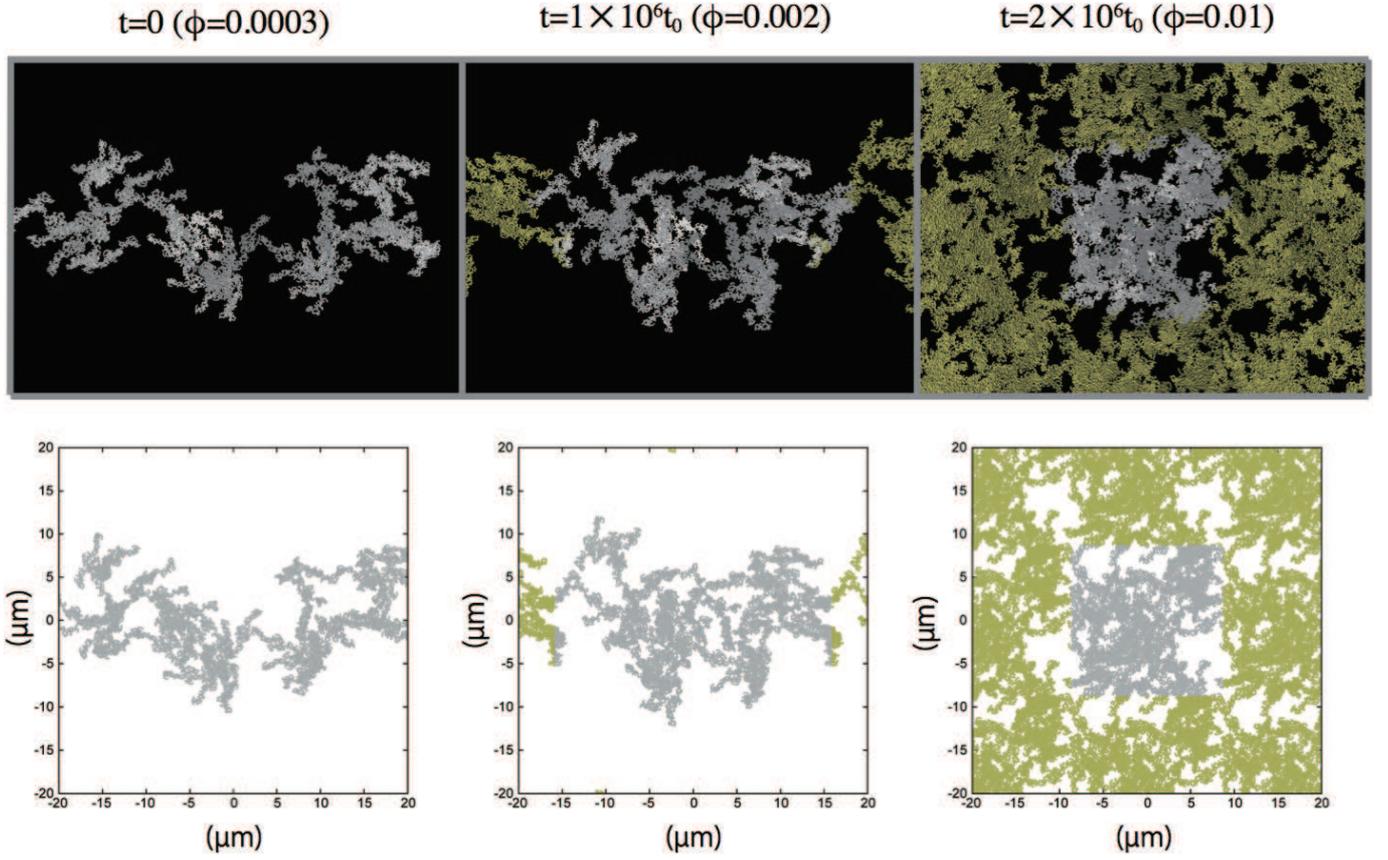}
 \end{center}
 \caption{
 Snapshots of the evolution of an aggregate under compression in the case of $N=16384$.
 The top three figures are three dimensional visualization.
 They have the same scale with different time epoch.
 The white particles are inside a box enclosed by the periodic boundaries.
 The yellow particles are in neighboring boxes to the box of white particles.
 For visualization, we do not draw the copies in back and front side of the boundaries but only 8 copies of the white particles across the boundaries.
 Each bottom figure represents projected positions onto two-dimensional plane of all particles in each corresponding top figure.
 The gray points in the bottom figures correspond to the positions of the white particles in the top figures and the yellow points correspond to those of the yellow particles in the top figures.
 Scales are in ${\rm \mu m}$.
 }
 \label{fig:compression}
\end{figure*}

The top three panels of Figure ~\ref{fig:compression} show snapshots of the evolution of an aggregate under compression in the case where $N=16384$, $C_{\rm v}=3\times 10^{-7}$, $k_{\rm n}=0$, and $\xi_{\rm crit}=8{\rm ~\AA}$.
The top three panels have the same scale but different time epochs, which are $t$ = 0, $1\times 10^{6}t_{0}$, and $2\times 10^{6}t_{0}$, respectively.
The white particles are inside the computational region enclosed by the periodic boundaries, while the yellow particles are in the neighbor copy regions.
(For visualization, we do not draw particles in the front- and back-side copy regions.)
The bottom three panels represent the projected positions onto two-dimensional plane for the correspondent top three figures.
We confirm that the dust aggregate is compressed by their copies from all directions.
As the compression proceeds, the aggregate of white particles is compressed by the neighboring aggregate of yellow particles.
We focus on how high pressure is generated by quasi-static compression in numerical simulations.
Our numerical simulations have several parameters; the size of the initial BCCA cluster, the compression rate, the normal damping force, and the critical displacement (corresponds to the rolling energy).
We investigate the dependence of the pressure on these parameters, by performing a lot of runs with different parameter sets.
Although we assume ice aggregates in most runs, we also investigate cases of silicate aggregates to compare them with previous studies.

\subsection{Fiducial run: obtaining the compression strength}\label{sec:shape}
\begin{figure*}
 \begin{center}
  \includegraphics[width=160mm]{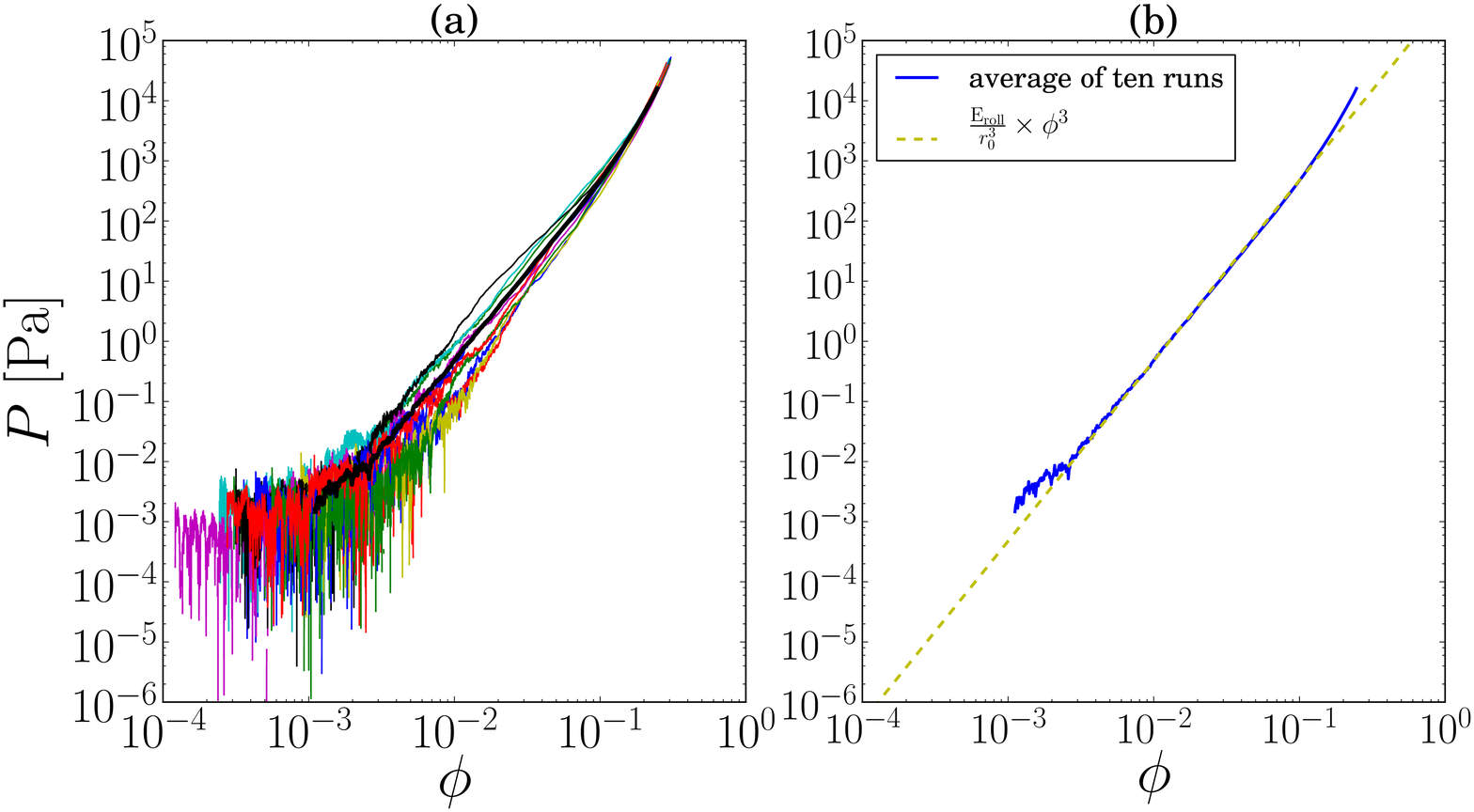}
 \end{center}
 \caption{
 (a) Pressure $P$ in [Pa] against filling factor $\phi$.
 The ten thin solid lines show the results for the initial BCCA clusters with different initial random numbers and thick solid line shows the arithmetic average of the ten runs.
 (b) Pressure $P$ in [Pa] against filling factor $\phi$. 
 Same as the thick solid line in (a) plotted with a dotted line of Equation (\ref{eq:eos}).
 The parameters are $N=16384$, $C_{\rm v}=3 \times 10^{-7}$, $k_{n}=0.01$, and $\xi_{\rm crit}=8 {\rm ~\AA}$.
 }
 \label{fig:randomnumber}
\end{figure*}

We put a BCCA cluster as the initial aggregate.
The BCCA cluster is created by sticking the copy of the aggregate from random direction.
The results depend on the random number of the initial condition, which is the shape of the BCCA aggregate.
To avoid the dependence, we take arithmetic averages of ten simulations of different initial conditions.
The pressure is measured using Equation (\ref{eq:pressure}) at each run.
We define the filling factor of an aggregate as
\begin{equation}
\phi = \frac{V_{0}N}{V},
\end{equation}
where $V_{0}$ is the monomer volume, $N$ is the number of monomers of the aggregate, and $V$ is the volume enclosed by the boundaries, which has a length of $L$.
The filling factor also can be written as $\phi=\rho/\rho_{0}$.
Figure  \ref{fig:randomnumber} shows that the measured pressure as a function of the filling factor $\phi(t)$.
The parameters of the simulations are $N=16384$, $C_{\rm v}=3 \times 10^{-7}$, $k_{\rm n}=0.01$, and $\xi_{\rm crit}=8 {\rm ~\AA}$.
The corresponding $E_{\rm roll}$ is $4.74\times 10^{-9} {\rm erg}$ for $\xi_{\rm crit}=8 {\rm ~\AA}$.
Each colored line in Figure \ref{fig:randomnumber}(a) shows each simulation with the different initial shape of the aggregate.
Figure ~\ref{fig:randomnumber}(b) shows the arithmetic average of the pressure measured in ten different runs.
Each line shows in different ranges of $\phi$.
The lowest $\phi$ is determined with the largest size of the initial boundary boxes of the ten runs.
We find that the compression strength is well reproduced by
\begin{equation}
P=P_{0}\phi^{3},
\end{equation}
where $P_{0}=4.74\times10^{5}$ Pa.
We analytically discuss why the compression strength is proportional to $\phi^{3}$ in Section \ref{sec:analytic}.
In the high density region ($\phi \gtrsim10^{-1}$), the measured strength deviates from the line of $P=P_{0}\phi^{3}$.
This is because the dissipation mechanism changes in the high density region (see Section \ref{sec:result_damp}).
The deviation in the low density region ($\phi \lesssim 3\times 10^{-3}$) is partly caused by a finite boundary speed (or compression rate) as discussed in the next subsection.
Another reason of the deviation in the low density region is related to the density of the initial BCCA cluster.
The filling factor of BCCA $\phi_{\rm BCCA}$ is estimated as,
\begin{equation}
\phi_{\rm BCCA} = \frac{V_{0}N}{V_{BCCA}}=\left(\frac{3}{5}\right)^{3/2}N^{-1/2},
\end{equation}
where we use the radius and the volume of a BCCA cluster, $r_{\rm BCCA}=\sqrt{5/3}N^{1/2}r_{0}$ and $V_{\rm BCCA}=(4 \pi/3)r_{\rm BCCA}^{3}$, respectively \citep[e.g.,][]{Suyama08}.
For $N=16384$, we obtain $\phi_{\rm BCCA} \sim 3 \times 10^{-3}$.
In the early stage of compression, $\phi$ is lower than $\phi_{\rm BCCA}$ because the initial BCCA clusters are apart from each other.
This space between BCCA clusters would also cause the deviation from the line of $P=P_{0}\phi^{3}$.

Now, we discuss the coefficient $P_{0}$ of the compression strength.
\citet{Wada08} shows that $E_{\rm roll}$ is important in the collisional compression strength.
Thus, $E_{\rm roll}$ is expected to be also important in the static compression strength.
Considering that the characteristic volume is monomer's volume $\sim r_{0}^{3}$, we suppose $P_{0}=E_{\rm roll}/r_{0}^{3}$, based on dimension analysis.
Therefore, the compression strength can be written as
\begin{equation}\label{eq:eos}
P=\frac{E_{\rm roll}}{r_{0}^3} \phi^3.
\end{equation}
We analytically discuss and confirm this equation in Section \ref{sec:analytic}.
We also plot this equation in Figure \ref{fig:randomnumber}(b).
This figure clearly shows that the result is well fitted by Equation (\ref{eq:eos}).

We show that compression strength is proportional to $\xi_{\rm crit}$, that is proportional to the rolling energy $E_{\rm roll}$ in Section \ref{sec:xi}.
We also confirm that Equation (\ref{eq:eos}) is applicable to the case of different $r_{0}$ in the silicate case.

\subsection{Dependence on the boundary speed}\label{sec:boundaryvelocity}
\begin{figure}[htbp]
 \begin{center}
  \includegraphics[width=80mm]{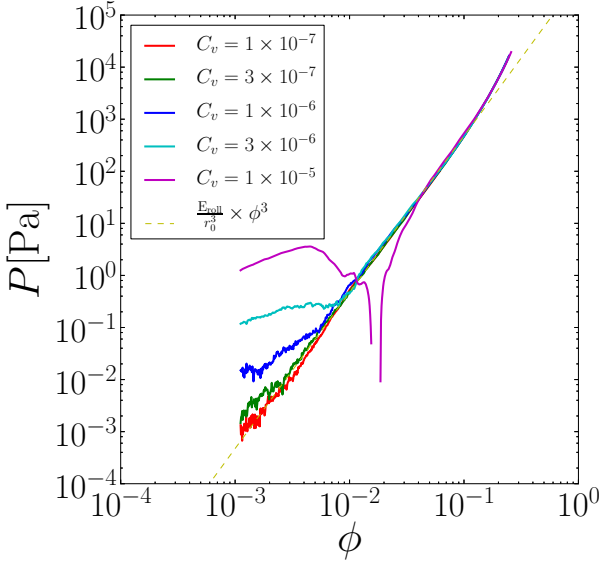}
 \end{center}
 \caption{
 Pressure $P$ in [Pa] against filling factor $\phi$ with different strain rate parameter $C_{\rm v}$.
 Each line shows the average of ten runs of the fixed strain rate: $C_{\rm v}=1\times 10^{-7},3\times 10^{-7},1\times 10^{-6},3\times 10^{-6},1\times 10^{-5}$.
 The other parameters are the same for every ten runs : $N=16384$, $k_{n}=0.01$, and $\xi_{\rm crit}=8 {\rm ~\AA}$.
 The dashed line is Equation (\ref{eq:eos}).
}
 \label{fig:velocity}
\end{figure}

To statically compress the aggregate, we should move the boundary at a sufficiently low velocity not to create inhomogeneous structure.
Figure \ref{fig:velocity} shows the dependency on the strain rate parameter.
Each line shows the average of ten runs.
The fixed parameters are $N=16384$, $k_{n}=0.01$, and $\xi_{\rm crit}=8 {\rm ~\AA}$.
The strain rate parameter $C_{\rm v}$ is equal to $1\times 10^{-7}$, $3\times 10^{-7}$, $1\times 10^{-6}$, $3\times 10^{-6}$, and $1\times 10^{-5}$, respectively.
The higher $C_{\rm v}$, the higher pressure in the low density region is required for compression.
This is mainly caused by the ram pressure from the boundaries with high speed.

When the compression proceeds and the density becomes higher to reach the line of Equation (\ref{eq:eos}), the pressure follows the equation.
From Figure \ref{fig:velocity}, $C_{\rm v}=3\times 10^{-7}$ creates sufficiently low boundary speed.
The boundary speed can be calculated as a function of $\phi$.
Using Equation (\ref{eq:vb}) and $\phi=(4/3)\pi r_{0}^{3}N/L^{3}$, the velocity difference between a boundary and the next boundary, $v_{\rm d}$, can be written as
\begin{equation}
v_{\rm d}=|2v_{\rm b}|=2\frac{C_{\rm v}}{t_{0}} \left( \frac{\frac{4}{3}\pi r_{0}^{3}N}{\phi}\right)^{1/3}
\end{equation}
In the case of $C_{\rm v}=3 \times 10^{-7}$, $v_{\rm d}$ = 12.7, 5.9, and 2.7 cm/s for $\phi$ = $10^{-3}$, $10^{-2}$, and $10^{-1}$, respectively.

Here, we discuss the velocity difference of boundaries, comparing with the effective sound speed of the aggregates.
The effective sound speed can be estimated as
\begin{equation}
c_{\rm s, eff}\sim \sqrt{\frac{P}{\rho}} \sim \sqrt{\frac{E_{\rm roll}}{\rho_{0}r_{0}^3}} \frac{\rho}{\rho_{0}} \sim \sqrt{\frac{E_{\rm roll}}{m_{0}}} \phi.
\end{equation}
where we use Equation (\ref{eq:eos}).
Using the rolling energy of ice particles, $c_{\rm s,eff}$ is given by
\begin{equation}
\label{eq:cs_ice}
c_{\rm s, eff} \sim 1.1\times10^{3} \phi ~{\rm cm/s}.
\end{equation}
Therefore, in the case of $C_{\rm v}=3\times 10^{-7}$, $v_{d}$ is not sufficiently low in the beginning of the simulation, where the aggregate has a low filling factor.
However, the boundary velocity difference reaches lower than the effective sound speed when $\phi \gtrsim 10^{-2}$.
 
\subsection{Dependence on the size of the initial BCCA cluster}\label{sec:size}
\begin{figure}[htbp]
 \begin{center}
  \includegraphics[width=80mm]{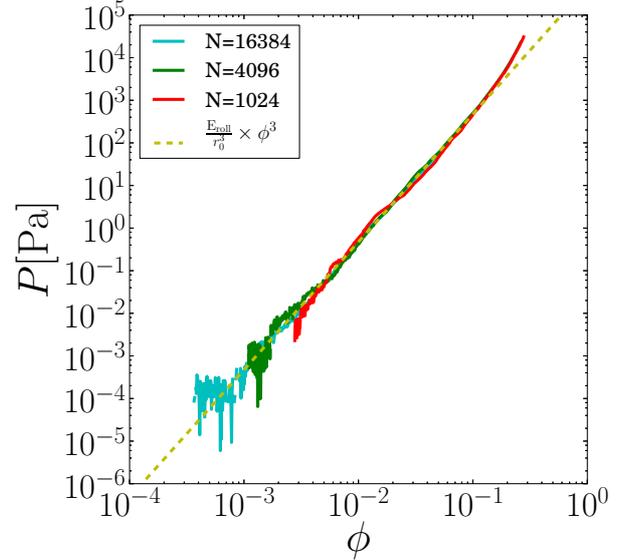}
 \end{center}
 \caption{
 Pressure $P$ in [Pa] against filling factor $\phi$ with different number of particles $N$.
 Each line shows the average of ten runs of the fixed number of particles: $N=1024, 4096,$ and $16384$.
 The other parameters are $C_{\rm v}=3 \times 10^{-7}$, $k_{\rm n}=0.01$, and $\xi_{\rm crit}=8 {\rm ~\AA}$ in the case of $N=1024, 4096$, and $C_{\rm v}=1 \times 10^{-7}$, $k_{\rm n}=0.01$, and $\xi_{\rm crit}=8 {\rm ~\AA}$ in the case of $N=16384$.
 The dashed line is Equation (\ref{eq:eos}).
  }
 \label{fig:particlenumber}
\end{figure}

To confirm that Equation (\ref{eq:eos}) is valid in the lower density region, we perform the simulations with the different number of particles, which is equivalent to the different sizes of the initial dust aggregates.
Figure \ref{fig:particlenumber} shows dependence on the number of particles of the initial BCCA cluster.
The initial numbers of particles are 1024, 4096, and 16384.
The other parameters are $C_{\rm v}=3 \times 10^{-7}$, $k_{\rm n}=0.01$, and $\xi_{\rm crit}=8 {\rm ~\AA}$ in the case of $N=1024$ and $N=4096$, and $C_{\rm v}=1 \times 10^{-7}$, $k_{\rm n}=0.01$, and $\xi_{\rm crit}=8 {\rm ~\AA}$ in the case of $N=16384$.
We chose lower $C_{\rm v}$ in the case of $N=16384$ in order to investigate the strength in lower $\phi$ region.
Each line represents the average of ten runs for each simulation as in Figures \ref{fig:randomnumber}(b) and \ref{fig:velocity}.
We draw the averaged line from the lower $\phi$ than that in Figure \ref{fig:velocity}.
In such a low $\phi$ region, we consider for some runs that the pressure is zero because the aggregate is isolated from the copies of the aggregate over the periodic boundaries.
Except for the initial deviation in low $\phi$, all lines have a good agreement with Equation (\ref{eq:eos}) where $\phi \lesssim 0.1$.
The result has the good agreement in lower $\phi$ for runs with larger $N$.
Therefore, we conclude that the formula Equation (\ref{eq:eos}) is valid for $\phi \lesssim 0.1$.

\subsection{Dependence on the normal damping force}\label{sec:result_damp}
\begin{figure}[htbp]
 \begin{center}
  \includegraphics[width=80mm]{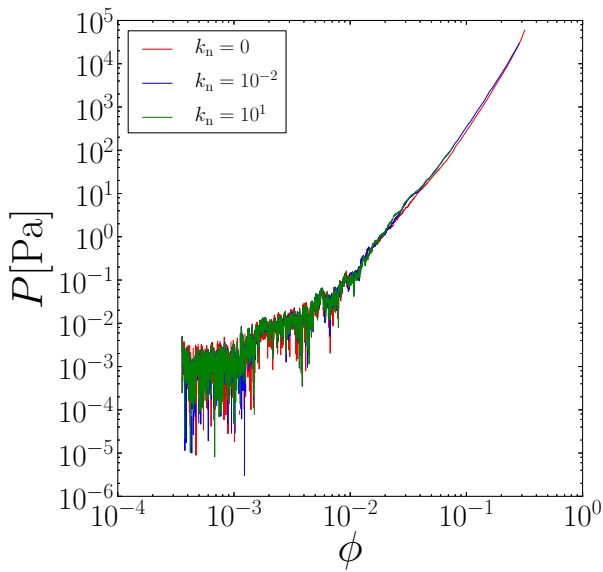}
 \end{center}
 \caption{
Pressure $P$ in [Pa] against filling factor $\phi$ with different normal damping force.
We put the same ten initial conditions varying the normal damping force with $k_{\rm n}=0$, $k_{\rm n}=10^{-2}$, and $k_{\rm n}=10^1$.
Each line shows the result of one run.
The other parameters are $N=16384$, $C_{\rm v}=3 \times 10^{-7}$, and $\xi_{\rm crit}=8 {\rm ~\AA}$.
 }
 \label{fig:damp}
\end{figure}

As described in Section \ref{sec:damp}, we adopt the normal damping force to reduce the normal oscillations in addition to \citet{Wada07}.
To confirm that this damping factor does not affect the simulation results, we set the damping factor $k_{\rm n}$ as a parameter.
Figure \ref{fig:damp} shows dependence of pressure on the normal damping factor $k_{\rm n}$.
The fixed parameters are $N=16384$ $C_{\rm v}=3\times 10^{-7}$, and $\xi_{\rm crit}=8 {\rm ~\AA}$.
Each line represents the result of one run for $k_{\rm n}= 0,10^{-2}$, and $10^{1}$, respectively.
This figure clearly shows that the normal damping force does not affect the simulation results.

\begin{figure}[htbp]
 \begin{center}
  \includegraphics[width=80mm]{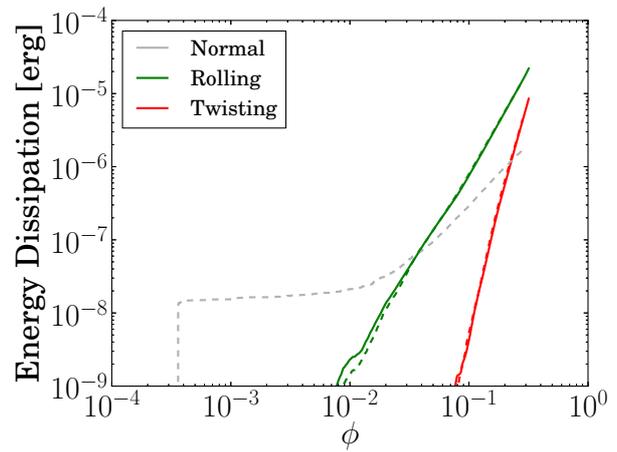}
 \end{center}
 \caption{
 Energy dissipation of each dissipation mechanism in [erg] against filling factor $\phi$.
 The solid lines shows the result in the case without the normal damping and the dashed lines in the case of $k_{\rm n}=0.01$ and
 The results in the case of $k_{\rm n}=10$ are not plotted because they are the same as those in the case of $k_{\rm n}=0.01$ and indistinguishable.
 The dissipation mechanisms are normal damping, rolling, sliding and twisting.
 The dissipation energy by sliding motion is less than $10^{-9}$ erg, 
 }
 \label{fig:dissipation}
\end{figure}

As mentioned in Section \ref{sec:shape}, the compression strength in the low density region ($\phi \lesssim 0.1$) is expected to be determined by the rolling motion.
In order to confirm this, we calculate the total energy dissipations of all motions, which are normal damping, rolling, sliding and twisting.
Figure \ref{fig:dissipation} shows the dissipated energy for each mechanism.
The solid lines represent the dissipated energies in the case without the normal damping and the dashed lines represent those in the case of $k_{\rm n}=0.01$.
The dissipated energy in the case of  $k_{\rm n}=10$ is indistinguishable from those in the case of $k_{\rm n}=0.01$, and thus we do not plot them.
Note that the dissipation energy of the sliding force is less than $10^{-9}$ erg, and thus it is not depicted in this figure.
The dissipation by the rolling and twisting is almost the same in the cases with and without the normal damping.
Thus, we confirm that the normal damping does not affect the compression strength although it dissipates the energy of the normal oscillations.
Aside from the normal dissipation, the dominant dissipation mechanism is the rolling motion.
This clearly shows that the static compression is determined by rolling motion of each connection, as mentioned in Section \ref{sec:shape}.
Where $\phi \gtrsim 0.1$, the energy dissipation by twisting motion occurs.
This is why Equation (\ref{eq:eos}) is valid until the filling factor reaches $0.1$ as mentioned in Section \ref{sec:shape}.
In the high density region, where $\phi \gtrsim 0.1$, another formulation is required but that is beyond the scope of this paper.

\subsection{Dependence on the rolling energy}\label{sec:xi}
\begin{figure}[htbp]
 \begin{center}
  \includegraphics[width=80mm]{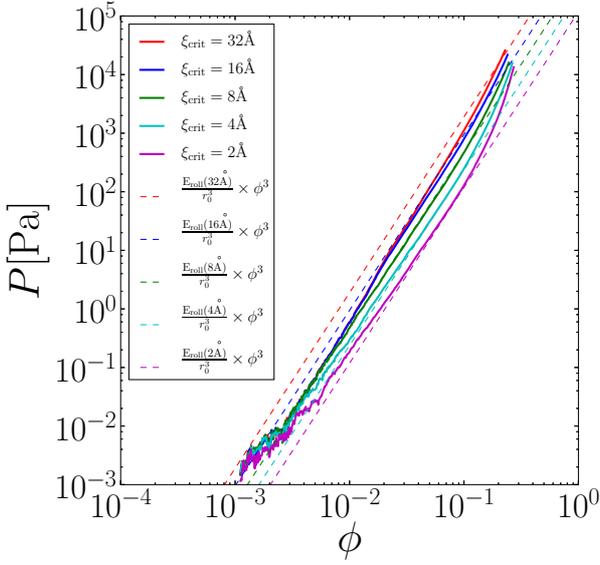}
 \end{center}
 \caption{
Pressure $P$ in [Pa] against filling factor $\phi$ with different critical displacement, $\xi_{\rm crit}$.
We put the same ten initial conditions varying $\xi_{\rm crit}$ with $\xi_{\rm crit}= 32, 16, 8, 4,$ and 2 \AA, respectively.
Each line shows the average of ten runs.
The other parameters are $N=16384$ $C_{\rm v}=3\times 10^{-7}$, and $k_{\rm n}= 10^{-2}$.
 }
 \label{fig:xi}
\end{figure}

\begin{figure}[htbp]
 \begin{center}
  \includegraphics[width=80mm]{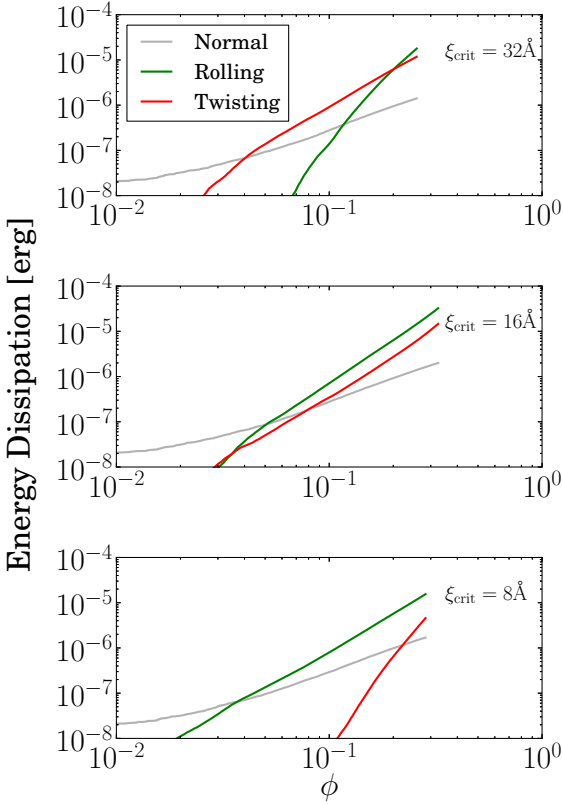}
 \end{center}
 \caption{
 Energy dissipation of each dissipation mechanism in [erg] against filling factor $\phi$.
 Each panel represents the case of different $\xi_{\rm crit}$, which are 32, 16, and 8 \AA, correspondent to Figure \ref{fig:xi}.
 }
 \label{fig:dissipation2}
\end{figure}

We also investigate the dependence of the compression strength on $\xi_{\rm crit}$.
Since $E_{\rm roll}$ is proportional to $\xi_{\rm crit}$, we investigate the dependence on the rolling energy in this section.
Figure \ref{fig:xi} shows that the dependency on $\xi_{\rm crit}$.
We vary $\xi_{\rm crit}$ with 32, 16, 8, 4, and 2 \AA.
The fixed parameters are $N=16384$, $C_{\rm v}=3\times 10^{-7}$, and $k_{\rm n}= 10^{-2}$.
This result shows that the compression strength is almost the same in the low density region.
This is because the periodic boundary creates the additional voids as discussed in Section \ref{sec:shape} and thus we should not focus on the low density region.
The lines in the case of $\xi_{\rm crit}=2,4,$ and $8{\rm ~\AA}$ are on their corresponding lines of Equation (\ref{eq:eos}) where $\phi \lesssim 0.1$.
The line in the case of $\xi_{\rm crit}=16 {\rm ~\AA}$ has a little deviation and that in the case of $\xi_{\rm crit}=32 {\rm ~\AA}$ has a deviation from their corresponding lines of Equation (\ref{eq:eos}).
The reason why the lines in the case of $\xi_{\rm crit} = 16, 32{\rm ~\AA}$ deviate from the corresponding lines of Equation (\ref{eq:eos}) is that the dissipation energy is dominated not by rolling motion but by twisting motion as indicated in Figure \ref{fig:dissipation2}.
This figure shows that dissipated energy of each dissipation mechanism.
We show the results of the cases with $\xi_{\rm crit}=$ 8, 16, and 32 \AA.
The normal damping is not contribute to the compression strength as discussed in Section \ref{sec:result_damp}, and thus we focus on the rolling and twisting motions.

When $\xi_{\rm crit} \le 8 {\rm \AA}$, the dissipation energy is dominated by rolling motion.
In the case of $\xi_{\rm crit} = 32 {\rm \AA}$, on the other hand, the dissipation energy is dominated by twisting motion.
In the case of $\xi_{\rm crit} = 16 {\rm \AA}$, the dissipation energy of rolling and twisting motion is comparable and thus this is the marginal case.
Thus, the reason why Equation (\ref{eq:eos}) is not valid when $\xi_{\rm crit} \ge 16 {\rm \AA}$ is that the twisting motion is the dominant mechanism to determine the compression strength.
Therefore, we conclude that Equation (\ref{eq:eos}) is valid when $\xi_{\rm crit} \le 8 {\rm \AA}$.

\subsection{Fractal structure}\label{sec:fractal}
\begin{figure}[htbp]
 \begin{center}
  \includegraphics[width=80mm]{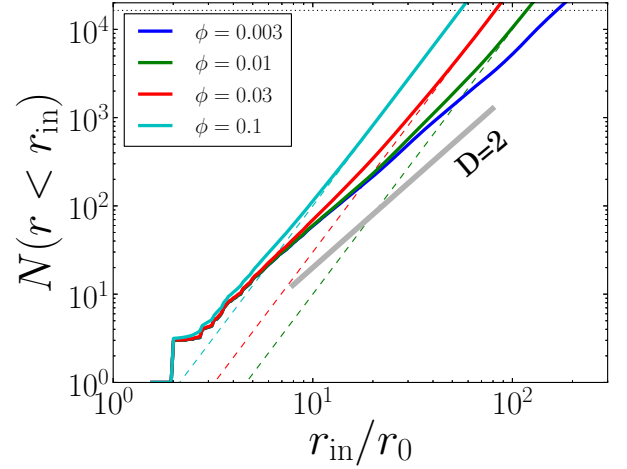}
 \end{center}
 \caption{
 Number of particles inside the radius $r_{\rm in}$ against normalized radius $r_{\rm in}/r_{0}$.
 This figure represents the fractal structure of the compressed aggregates in our simulation for various $\phi$.
 We set a particle as an origin and count the number of particles inside $r < r_{\rm in}$, where $r$ is the distance from the origin to each particle's center.
 Then we count the same correlation of all particles as an origin and take their average (similar figure of Figure.7 in the paper of \citet{Wada08}).
 Each line shows the result at the different time step.
 The solid thick lines represents the structure of fractal dimension $D=2$, and dashed lines represent  $D=3$ for each corresponding $\phi$.
 The dotted line shows the number of particles in calculation.
 The region below this line corresponds to inside the periodic boundaries.
 }
 \label{fig:count}
\end{figure}

We also investigate how the fractal structure of the dust aggregate changes.
Figure \ref{fig:count} shows how many particles are inside the distance $r_{\rm in}$ for four snapshots.
We select one run from the case with $N=16384$, $C_{\rm v}=3\times 10^{-7}$, $k_{\rm n}= 10^{-2}$, and $\xi_{\rm crit}=8 {\rm ~\AA}$.
Each snapshot is when $\phi=0.003, 0.01, 0.03$ and 0.1, respectively.
We take a particle as an origin and count the number of particles inside $r < r_{\rm in}$, where $r$ is the length from the origin.
Then we set for all the other particles inside the computational region as an origin and take an average of them.
We obtained the same trend in several runs in the cases of different shapes of initial aggregates.

Note that we also count particles beyond the periodic boundaries.
In high $r_{\rm in}$, $N\propto r_{\rm in}^3$ because copies over the periodic boundary distributed as fractal dimension of 3.
Therefore, where $N(r<r_{\rm in}) \gtrsim 16384$, $N$ must be $N\propto r_{\rm in}^3$.
However, it is almost out of range of Figure \ref{fig:count}.
The dotted line in Figure \ref{fig:count} shows the number of particles in calculation, which is $N=16384$.
The results over this line is affected by the periodic boundary condition and those below this line is in computational region.
Thus, the results below the line represents the fractal structure inside the computational region and are not the artificial effect of the periodic boundary condition.

Since the initial aggregate is a BCCA cluster, $N$ is proportional to $r_{\rm in}^2$.
In the case of $\phi=0.003$, which is equivalent to $\phi$ of the initial BCCA cluster, $N\propto r_{\rm in}^2$ as shown in Figure \ref{fig:count}.
When the fractal dimension is $3$, $N$ can be written as
\begin{equation}
N(r< r_{\rm in})=\frac{\phi V(r<r_{\rm in})}{V_{0}}=\phi \left(\frac{r_{\rm in}}{r_{0}}\right)^{3},
\end{equation}
where $V(r<r_{\rm in})=(4/3)\pi r_{\rm in}^{3}$.
We also plot this equation as dashed lines for each $\phi$ in Figure \ref{fig:count}.
Each dashed line has a good agreement in the large scale, while maintaining $N\propto r_{\rm in}^2$ in small scale.

Therefore, the structure evolution in the static compression is as follows.
Initially, $N\propto r_{\rm in}^2$ because the aggregate is a BCCA cluster.
As compression proceeds, the fractal dimension $D$ becomes 3 in a large scale while it is 2 in a small scale.
The transit scale from $D=2$ to $D=3$ becomes smaller as compression proceeds until $D=3$ in any scale.
This structure evolution means that the static compression reconstructs the aggregate first in a large scale with keeping the small scale BCCA structure.
This is the reason why the rolling motion determines the compression strength, as discussed in Section \ref{sec:analytic}.

\subsection{Silicate case : Comparison with previous studies}
\begin{figure*}[htbp]
 \begin{center}
  \includegraphics[width=160mm]{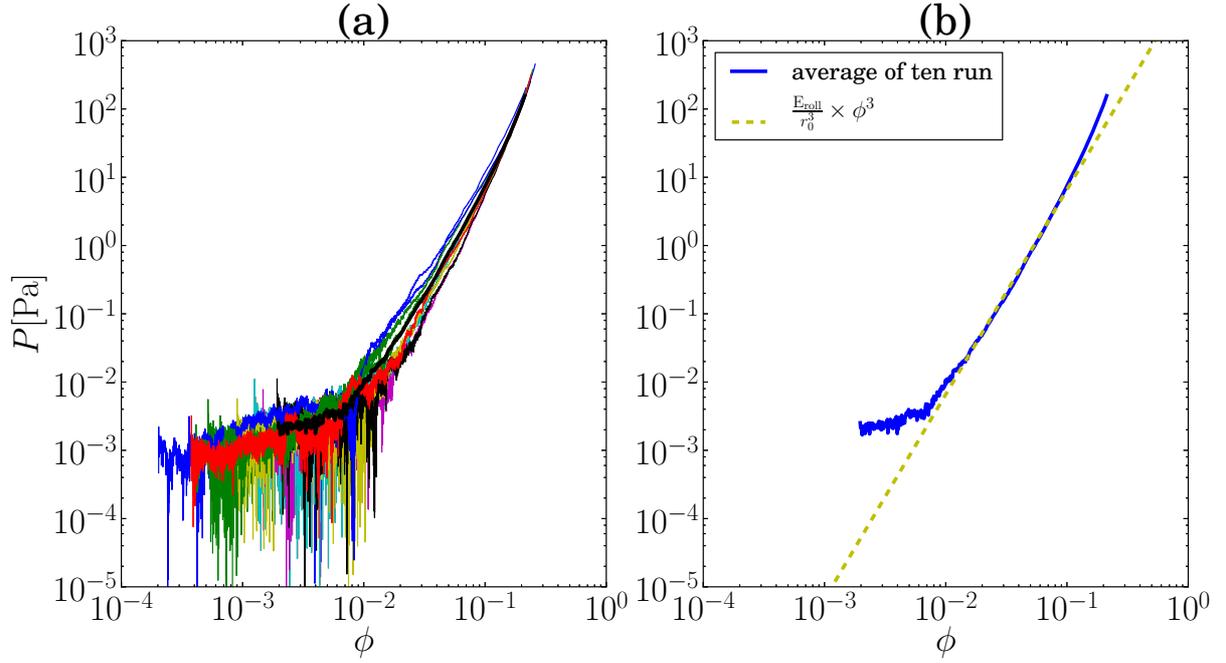}
 \end{center}
 \caption{
 Pressure $P$ in [Pa] against filling factor $\phi$.
 This figure is same as Figure \ref{fig:randomnumber} but for the case of silicate particles ($r_{0}=0.6 {\rm \mu m}$).
 }
 \label{fig:sili}
\end{figure*}

The compression strength has been investigated in the previous study \citep{Seizinger12}.
To investigate the connection of compression strength from the low density to the high density region, we perform simulations in the case of silicate with the same parameters of \citet{Seizinger12}.
Figure \ref{fig:sili} shows compression in the case of silicate whose monomer size is $0.6 {\rm ~\mu m}$.
The parameters are $N=16384$, $C_{\rm v}=3\times 10^{-7}$, and $k_{\rm n}=0.01$.
The solid lines in Figure \ref{fig:sili}(a) show the results of ten runs with different initial aggregates and the thick solid line in Figure \ref{fig:sili}(b) shows their average.
Using the rolling energy of silicate, which is $E_{\rm roll}=1.42 \times 10^{-8}$ erg, we also plot the line of Equation (\ref{eq:eos}) in Figure \ref{fig:sili}(b).
Since $t_{0}$ is given by $1.71 \times 10^{-9}$ sec in the case of silicate aggregates, $v_{\rm d}$ becomes 4.01 cm/s for $\phi = 10^{-2}$ with $C_{\rm v}$=$3\times 10^{-7}$.
This $v_{\rm d}$ is larger than $c_{\rm s, eff}$ (= 0.77 cm/s when $\phi=0.01$) for silicate aggregates, allowing the numerical results shown in Figure \ref{fig:sili} to deviate from the line of Equation (\ref{eq:eos}) in the low $\phi$ region.
When $v_{\rm d}=c_{\rm s, eff}$, $\phi=3.4 \times 10^{-2}$, and therefore, the compression strength should obey Equation (\ref{eq:eos}) when $\phi \gtrsim 3.4 \times 10^{-2}$.
In the case of silicate, computational time is huge compared with ice particle cases.
We take relatively high value of the boundary speed to save the computational time.
Therefore, the result is deviate from Equation (\ref{eq:eos}) in the low density region because of the high velocity.
In other words, the compression is not static in the low density region.
In the high density region, on the other hand, the result is in good agreement with Equation (\ref{eq:eos}), suggesting that Equation (\ref{eq:eos}) is applicable to aggregates consisting of silicate particles with different $r_{0}$.

\begin{figure}[htbp]
 \begin{center}
  \includegraphics[width=80mm]{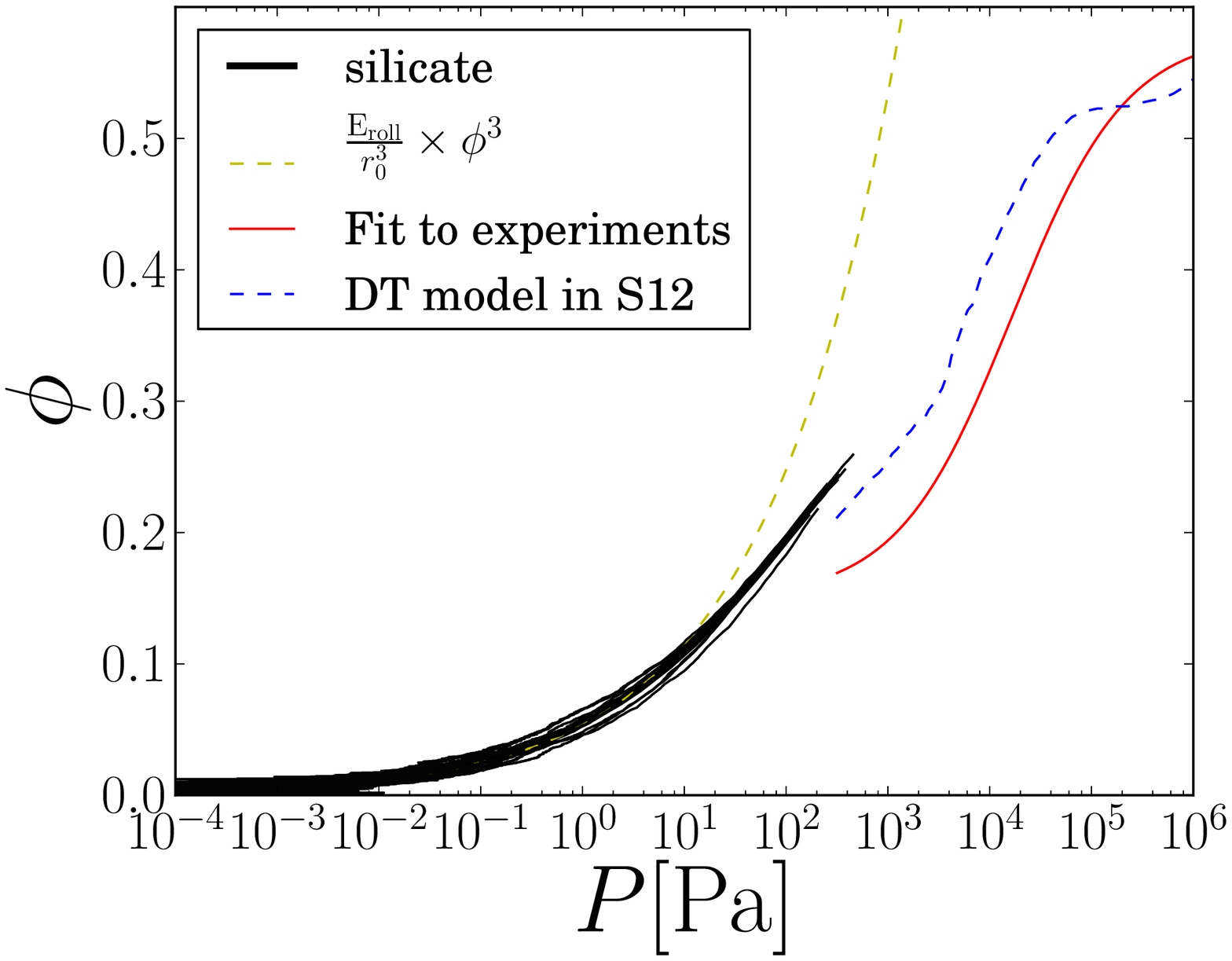}
 \end{center}
 \caption{
 The filling factor $\phi$ against pressure $P$ in [Pa].
 This figure is same as Figure \ref{fig:sili}, but plotted with linear scale of $\phi$ and reversal of $xy$ axis to compare with previous studies (see Figure 4 in \citet{Seizinger12}).
 The dotted line is the result of numerical simulations in the high density region ($\phi\gtrsim0.1$) in \citet{Seizinger12} and the thin solid line is the fitting formula proposed by \citet{Guttler09}.
 Our results consistently connect to the previous simulations in the high density region.
 }
 \label{fig:compare}
\end{figure}

To directly compare with previous studies, Figure \ref{fig:compare} shows the filling factor in linear scale against pressure in log scale.
This figure corresponds to Figure 4 in \citet{Seizinger12}.
The solid lines are our simulation results and the dashed line is Equation (\ref{eq:eos}) in the low density region ($\phi < 0.1$).
The dotted and solid lines are the result of \citet{Seizinger12} and the fitting formula to experiments \citep{Guttler09}, respectively.
They performed similar $N$-body simulations to ours but using a BPCA aggregate composed of silicate particles as an initial condition.
The compression strength of our simulations has a good agreement with the same interaction model in \citet{Seizinger12} with a little discrepancy: $\phi=0.24$ at $P=300$ Pa in our simulations and $\phi=0.21$ at $P=300$ Pa in \citet{Seizinger12}.
The discrepancy, 13\% in $\phi$ may be caused by the difference in the initial aggregate or the pressure measurement method.
The fitting formula of \citet{Guttler09} suggests $\phi=0.17$ at $P=300$ in the experiments.
The discrepancy from our simulations is 29 \% in $\phi$.
In applicable uses of the static compression formula, we focus on obtaining $\phi$ with a given $P$.

\section{Understanding the compression strength formula}
\label{sec:analytic}
\begin{figure*}[htbp]
 \begin{center}
  \includegraphics[width=160mm]{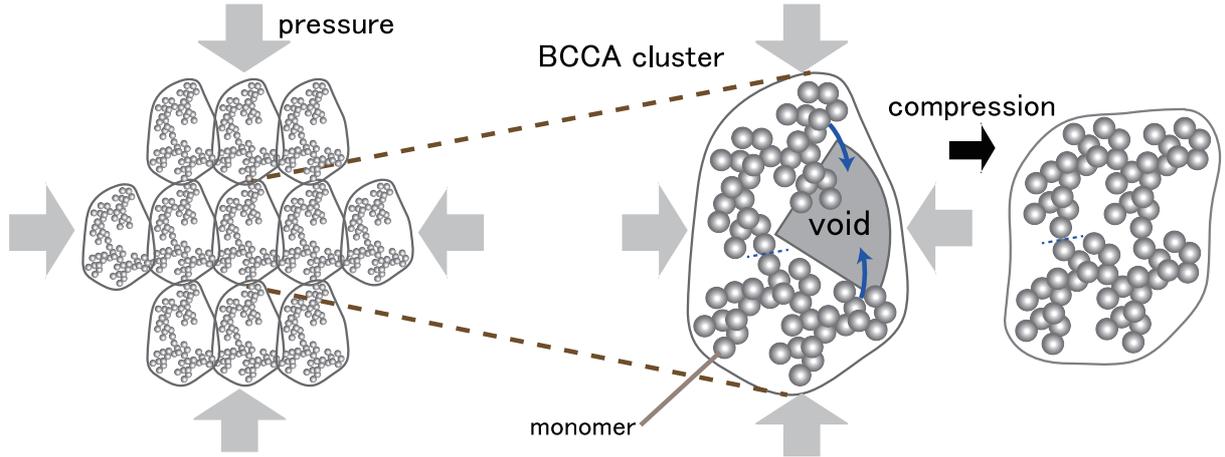}
 \end{center}
 \caption{
 Schematic drawing of compression of a dust aggregate consisting of a number of BCCA clusters.
 The left figure shows a dust aggregate consisting of many BCCA clusters and the BCCA clusters are distributed three dimensionally.
 Each enclosed line represents each region dominated by the BCCA clusters.
 The central figure is a BCCA cluster, receiving pressure from the next clusters.
 The BCCA cluster has a large void depicted in the central figure, and thus the void would be compressed, as expressed in the right figure.
 The required energy to compress the void is the energy to rotate the connection of monomers in contact.
 Therefore, the compression can be determined by the rolling motion of monomer connection on the connecting point of the subclusters.
 }
 \label{fig:rolling}
\end{figure*}

In this section, we analytically derive the compression strength and confirm Equation (\ref{eq:eos}).
First, we consider the structure of a fluffy aggregate in static compression in our simulations.
As described in Section \ref{sec:periodic}, we adopt the periodic boundary condition and put a BCCA cluster as the initial condition.
This corresponds to a large aggregate which filled up with BCCA clusters three dimensionally.
As compression proceeds, the initial BCCA cluster is compressed but the aggregate keeps smaller BCCA structure as confirmed in Section \ref{sec:fractal}.
Therefore, the aggregate in static compression always consists of BCCA clusters in some scale and filled up with them.
Figure~\ref{fig:rolling} illustrates the aggregate in static compression.
The enclosed lines depict BCCA clusters in a small scale.

Next, we consider why the compression strength can be determined by the rolling energy.
The internal mass density and the volume filling factor of the aggregate are equal to those of the BCCA clusters.
Compression of the whole aggregate proceeds by compression of each cluster.
Therefore, the compression strength of the whole aggregate would be determined by BCCA clusters.
The right panel of Figure \ref{fig:rolling} illustrates compression of one of BCCA clusters.
The pressure on the BCCA cluster is exerted by neighbor clusters, which causes the compression of the BCCA cluster. 
The BCCA cluster can be further divided into two smaller subclusters because BCCA clusters are created by cluster-cluster aggregation. 
A large void exists between the two smaller clusters and they are connected with one connection of monomers in contact, represented by dashed line in the right panel of Figure \ref{fig:rolling}.
The compression of the BCCA cluster occurs by crashing the large void, which requires only rolling of the monomers at the connection.

Now, let us estimate the compression strength.
In static compression, the aggregate is compressed by external pressure.
Each BCCA cluster feels a similar pressure, $P$.
Using the pressure, the force on the BCCA cluster is approximately given by
\begin{equation}
\label{eq:F}
F \sim P\cdot r_{\rm BCCA}^2.
\end{equation}
Since the crashing of the large void is accompanied by rolling of a pair of monomers in contact, the work required for the crashing is given by so-called the rolling energy of monomers, $E_{\rm roll}$ (\citet{DominikTielens97} or see Equation (\ref{eq:Eroll}) for its definition).
Therefore, the required force to compress the aggregate satisfies,
\begin{equation}
F \cdot r_{\rm BCCA} \sim E_{\rm roll}.
\end{equation}
Substituting Equation (\ref{eq:F}), we further obtain the required pressure to compress the aggregate as
\begin{equation}
\label{eq:P}
P \sim\frac{E_{\rm roll}}{r_{\rm BCCA}^3}.
\end{equation}

The radius of the BCCA clusters can be written by using the physical values of the whole aggregate.
The internal density of the BCCA cluster is dependent on its radius.
The BCCA cluster has the fractal dimension of 2, and its radius is approximately given by $r_{\rm BCCA} = N^{1/2}r_{0}$, where $N$ is the number of constituent monomers in the BCCA subcluster.
The internal density of the BCCA cluster is evaluated as
\begin{equation}
\label{eq:rhobcca}
\rho \sim \frac{N m_{0}}{r_{\rm BCCA}^3} \sim \left(\frac{r_{\rm BCCA}}{r_{0}}\right)^{-1} \rho_{0}.
\end{equation}
Using equations (\ref{eq:P}) and (\ref{eq:rhobcca}), we finally obtain the required pressure (or the compression strength) as
\begin{equation}
P\sim\frac{E_{\rm roll}}{r_{0}^3} \left(\frac{\rho}{\rho_0}\right)^3.
\end{equation}
This is the same as Equation (\ref{eq:eos}) obtained from our numerical simulations.

\section{Summary}\label{sec:summary}
We investigated the static compression strength of highly porous dust aggregates, whose filling factor $\phi$ is lower than 0.1.
We performed numerical $N$-body simulations of static compression of highly porous dust aggregates.
The initial dust aggregate is assumed to be a BCCA cluster.
The particle-particle interaction model is based on \citet{DominikTielens97} and \citet{Wada07}.
We introduced a new method for compression.
We adopted the periodic boundary condition in order to compress the dust aggregate uniformly and naturally.
Because of the periodic boundary condition, the dust aggregate in computational region represents a part of a large aggregate, and thus we could investigate the compression of a large aggregate.
The periodic boundaries move toward the center and the distance between the boundaries becomes small.
To measure the pressure of the aggregate, we adopted a similar manner used in molecular dynamics simulations.
As a result of the numerical simulations, our main findings are as follows.

\begin{itemize}

\item
The relation between the compression pressure $P$ and the filling factor $\phi$ can be written as
\begin{equation}
\label{eq:eos_final}
P = \frac{E_{\rm roll}}{r_{0}^3}\phi^3,
\end{equation}
where $E_{\rm roll}$ is the rolling energy of monomer particles and $r_{0}$ is the monomer radius.
We defined the filling factor as $\phi=\rho / \rho_{0}$, where $\rho$ is the mass density of the whole aggregate, and $\rho_{0}$ is the material mass density.
Equation (\ref{eq:eos_final}) is independent of the numerical parameters; the number of particles, the  size of the initial BCCA cluster, the boundary speed, the normal damping force.
We confirmed that Equation (\ref{eq:eos_final}) is applicable in different $E_{\rm roll}$ and $r_{0}$.
We also analytically confirmed Equation (\ref{eq:eos_final}).
\item 
Equation (\ref{eq:eos_final}) is valid where $\phi \lesssim 0.1$ in the high density region.
In the low density region, we confirmed that Equation (\ref{eq:eos_final}) is valid for $\phi \gtrsim 10^{-3}$ in the case of $N=16384$.
From the results of different initial sizes of the aggregates, Equation (\ref{eq:eos_final}) is valid in the lower density region in the case of the larger aggregates.
\item 
The initial BCCA cluster has a fractal dimension of 2 in the radius of the cluster, although the whole aggregate has a fractal dimension of 3 because of the periodic boundary.
As compression proceeds, the fractal dimension inside the radius of the initial BCCA cluster becomes 3, while the fractal dimension in smaller scale keeps being 2.
This means that the initial set up, which is that fractal dimension in large scale is 3 and that in small scale is 2, well reproduce the structure of a dust aggregate in static compression as a consequence.
This also supports the fact that the compression strength is determined by BCCA structure in a small scale.
\item
The static compression in the high density region ($\phi \gtrsim 0.1$) has been investigated in silicate case in previous studies \citep{Seizinger12}.
We performed the numerical simulations in silicate case and confirmed that our results are consistent with that of previous studies in the high density region.
\end{itemize}

The compression strength formula allows us to study how static compression affects the porosity evolution of dust aggregates in protoplanetary disks.
In application to dust compression in protoplanetary disks, we use the compression strength formula to obtain $\phi$ with a given $P$.
Moreover, the obtained compression strength would be applicable to SPH simulations of dust collisions.
Such application of the static compression process is important future work.
In this work, we did not study shear or tensile strengths.
Those strengths are worth investigated in future work.

\begin{acknowledgements}
We thank Kohji Tomisaka and Hiroki Senshu for fruitful discussions.
We appreciate careful reading and fruitful comments by the anonymous referee and the editor, Tristan Guillot.
A.K. is supported by the Research Fellowship from the Japan Society for the Promotion of Science (JSPS) for Young Scientists (24$\cdot$2120).
S.O. acknowledges support by Grants-in- Aid for JSPS Fellows (22$\cdot$7006) from MEXT of Japan.
K.W. acknowledges support by Grants-in- Aid for Scientific Research (24540459) from MEXT of Japan.

\end{acknowledgements}

\bibliography{compression}

\end{document}